\documentclass[twocolumn,aps,superscriptaddress,pre,nofootinbib]{revtex4}
\usepackage{natbib}
\usepackage{amsmath,amssymb,graphicx}
\usepackage{algorithmic}
\usepackage{enumerate}
\usepackage{times}
\usepackage{color}
\usepackage{soul}
\usepackage{textcomp}
\usepackage[pdftex]{hyperref}
\hypersetup{colorlinks=true,linkcolor=blue,citecolor=blue,urlcolor=blue}

\begin{document}

\title{Optimization of population annealing Monte Carlo for large-scale spin-glass simulations}

\author{Amin Barzegar}
\email{a.barzegar@physics.tamu.edu}
\affiliation{Department of Physics and Astronomy, Texas A\&M University,
College Station, Texas 77843-4242, USA}

\author{Christopher Pattison}
\email{cpattison@tamu.edu}
\affiliation{Department of Physics and Astronomy, Texas A\&M University,
College Station, Texas 77843-4242, USA}

\author{Wenlong Wang}
\email{wenlongcmp@gmail.com}
\affiliation{Department of Physics and Astronomy, Texas A\&M University,
College Station, Texas 77843-4242, USA}

\author{Helmut G. Katzgraber}
\affiliation{Microsoft Quantum, Microsoft, Redmond, WA 98052, USA}
\affiliation{Department of Physics and Astronomy, Texas A\&M University,
College Station, Texas 77843-4242, USA}
\affiliation{Santa Fe Institute, Santa Fe, New Mexico 87501 USA}

\begin{abstract}

Population annealing Monte Carlo is an efficient sequential algorithm
for simulating $k$-local Boolean Hamiltonians. Because of its structure,
the algorithm is inherently parallel and therefore well suited for
large-scale simulations of computationally hard problems. Here we
present various ways of optimizing population annealing Monte Carlo
using $2$-local spin-glass Hamiltonians as a case study. We demonstrate
how the algorithm can be optimized from an implementation, algorithmic
accelerator, as well as scalable parallelization points of view. This
makes population annealing Monte Carlo perfectly suited to study other
frustrated problems such as pyrochlore lattices, constraint-satisfaction
problems, as well as higher-order Hamiltonians commonly found in, e.g.,
topological color codes.

\end{abstract}

\maketitle

\section{Introduction}

Monte Carlo algorithms are widely used in many areas of science,
engineering, and mathematics. These approaches are of paramount
importance for problems where no analytical solutions are possible. For
example, the class of Ising-like Hamiltonians can only be solved
analytically in few exceptionally rare cases. The vanilla Ising model
can only be solved analytically in one, two, as well as infinite space
dimensions. A solution in three space dimensions remains to be found to
date \cite{ising:25,huang:87}.  Therefore, simulations are necessary to
understand these systems in three space dimensions. The situation is far
more dire when more complex interactions---such as $k$-local terms
rather than the usual quadratic or $2$-local terms---are used.
Similarly, the inclusion of disorder allows for analytical solutions
only in the mean-field regime
\cite{edwards:75,parisi:79,sherrington:75,binder:86,stein:13}.  These
spin-glass problems, a subset of frustrated and glassy systems,
represent the easiest $2$-local Hamiltonian that is computationally
extremely hard. A combination of diverging algorithmic timescales (with
the size of the input) due to rough energy landscapes and the need for
configurational (disorder) averages to compute thermodynamic quantities
makes them the perfect benchmark problem to study novel algorithms.
Finally, computing ground states of spin glasses on nonplanar graphs is
an NP-hard problem where Monte Carlo methods have been known to be
efficient heuristics \cite{katzgraber:04c,wang:15,zhu:15b} and where
only few efficient exact methods exist for small system sizes.

It is therefore of much importance to design or improve efficient
algorithms either to save computational effort or have better quality
data with the same computational effort when studying these complex
systems. Two popular algorithms that are currently in use (for both
thermal sampling, as well as optimization) are parallel tempering (PT)
Monte Carlo \cite{geyer:91,hukushima:96} and population annealing Monte
Carlo (PAMC) \cite{hukushima:03,zhou:10,machta:10,wang:15e}. 

Although both PT and PAMC are extended ensemble Monte Carlo methods,
PAMC is a sequential Monte Carlo algorithm, in contrast to PT, which is
a (replica-exchange) Markov-chain Monte Carlo method. PAMC is a
population-based Monte Carlo method and thus well suited for
implementations on multicore high-performance computing machines.  PAMC
is similar to simulated annealing \cite{kirkpatrick:83}, however, with
an extra resampling step when the temperature is reduced to maintain
thermal equilibrium.  PT has been intensively optimized and has been to
date the work horse in statistical physics and is equally efficient
in simulating spin glasses when compared to PAMC \cite{wang:15e}.
PAMC, on the other hand, remains a relatively new simulation method.  Although
careful systematic studies of PAMC \cite{wang:14,wang:15e} exist, and
the method has been applied broadly
\cite{wang:15,borovsky:16,barash:17,callaham:17}, little
effort has been made to thoroughly optimize the algorithm.  Here we
focus on this problem and study various approaches to improve the
efficiency of PAMC for large-scale simulations. While some approaches
improve PAMC, others have little to no effect.  Note that related
optimization ideas are explored in Ref.~\cite{amey:18}.

\begin{figure}[htb]
\begin{center}
\includegraphics[width=1.03\columnwidth]{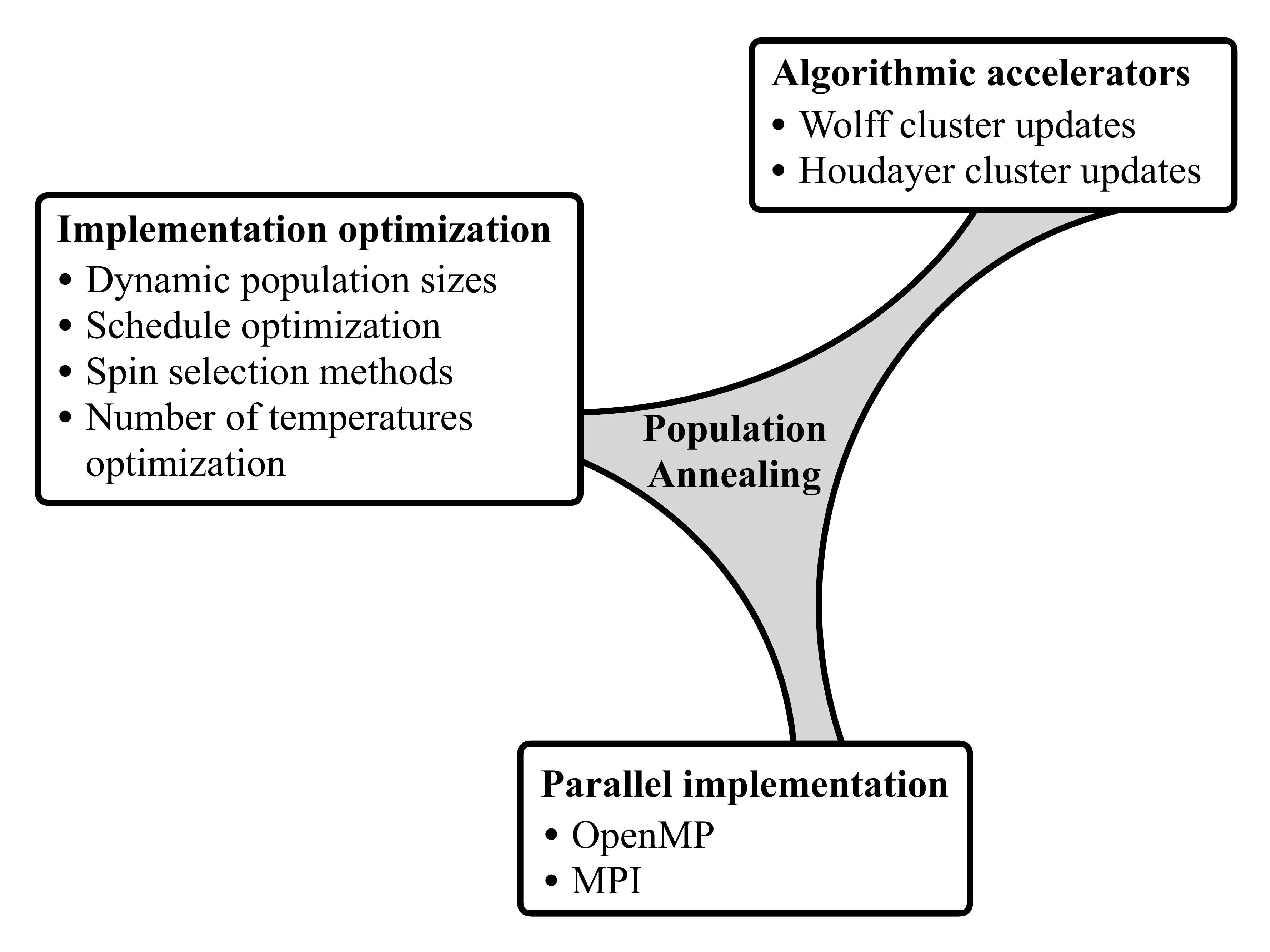}
\caption{
Diagram outlining the different optimizations we have implemented for
population annealing Monte Carlo. These range from optimizations in the
implementation, such as efficient spin selection techniques, to
algorithmic accelerators (e.g., the inclusion of cluster updates), as
well as parallel implementations. See the main text for details.
}
\label{DM}
\end{center}
\end{figure}

Our strategy to optimize PAMC is three pronged, as illustrated in
Fig.~\ref{DM}. First, we study different implementation optimizations.
Here we discuss dynamic population sizes that vary with the temperature
during the anneal, as well as the optimization of different annealing
schedules.  We also investigate different spin selection methods (order of
spin updates in the simulation) such as random, sequential and
checkerboard. While for disordered systems sequential updates are
commonplace, random updates are needed for nonequilibrium studies.  In
the case of bipartite lattices, a checkerboard spin-update technique can
be used, which is perfectly suited for parallelization.  Furthermore, we
discuss how to determine the optimum number of temperatures for a given
simulation. Second, we analyze the effects of algorithmic
accelerators by adding cluster updates to PAMC.  We have
studied Wolff cluster updates \cite{wolff:89}, as well as Houdayer
cluster updates \cite{houdayer:01}, and isoenergetic cluster moves
\cite{zhu:15b}.  Third, we discuss different parallel implementations
using both OpenMP (ideal for shared-memory machines
\cite{wang:14,wang:15e,comment:openmp}) and MPI \cite{comment:mpi} with
load balancing (ideal for scalable massively parallel implementations).
Note that PAMC implemented on graphics processing units has been
discussed extensively in Refs.~\cite{borovsky:16,barash:17a}.

The paper is structured as follows. We first introduce in Sec.~\ref{mm}
some concepts needed in this study, such as the case study Hamiltonian
and outline the PAMC algorithm. Implementation optimizations are
presented in Sec.~\ref{imp}, algorithmic accelerators via cluster
updates in Sec.~\ref{acc}, and parallel implementations are discussed in
Sec.~\ref{parallel}, followed by concluding remarks.

\section{Preliminaries}
\label{mm}

In this section we introduce some concepts needed for the PAMC
optimization in the subsequent section. In particular, we introduce the
Ising spin-glass Hamiltonian (our case study), as well as PAMC and
different algorithmic accelerators.

\subsection{Case study: Spin glasses}

We study the zero-field two-dimensional (2D) and 3D
Edwards-Anderson Ising spin glass \cite{edwards:75} given by the
Hamiltonian
\begin{equation}
H = - \sum_{\langle ij \rangle} J_{ij} S_i S_j ,
\label{eq:ham}
\end{equation}
where $S_i=\pm 1$ are Ising spins and the sum is over the nearest
neighbors on a $D$-dimensional lattice of linear size $L$ with $N_{\rm
spin} = L^{D}$ spins. The random couplings $J_{ij}$ are chosen from a
Gaussian distribution with mean zero and variance $1$. We refer to each
disorder realization as an ``instance.'' The model has no phase
transition to a spin-glass phase in 2D \cite{singh:86}, while in 3D
there is a spin-glass phase transition at $T_{\rm c} \approx 0.96$
\cite{katzgraber:06} for Gaussian disorder.

\subsection{Outline of population annealing Monte Carlo}

Population annealing Monte Carlo
\cite{hukushima:03,zhou:10,machta:10,wang:15e} is similar to simulated
annealing (SA) \cite{kirkpatrick:83} in many ways. For example, both
methods are sequential. However, the most important differentiating
aspect between PAMC and SA is the addition of a population of replicas
that are resampled when the temperature is lowered in the annealing
schedule.

PAMC \cite{wang:15e} starts with a large population of $R$ replicas at a
high temperature, where thermalization is easy. In our simulations, we
initialize replicas randomly at the inverse temperature $\beta = 1/T
=0$. The population traverses an annealing schedule with $N_T$
temperatures and maintains thermal equilibrium to a low target
temperature, $T_{\rm{min}} = 1/\beta_{\rm{max}}$. When the temperature
is lowered from $\beta$ to $\beta^\prime$, the population is resampled.
The mean number of the copies of replica $i$ is proportional to the
appropriate reweighting factor, $\exp[-(\beta^\prime-\beta) E_i]$. The
constant of proportionality is chosen such that the expectation value of
the population size at the new temperature is $R(\beta')$. Note that
$R(\beta')$ is usually kept close to $R$; however, this is not a
necessary condition.  Indeed, in our dynamical population size
implementation, we let $R$ change as a function of $\beta$ and seek
better algorithmic efficiency in the number of spin updates. The
resampling is followed by $N_{\rm S}=10$ Monte Carlo sweeps (one Monte
Carlo sweep represents $N_{\rm spin}$ attempted spin updates) for each
replica of the new population using the Metropolis algorithm. We keep
$N_{\rm S}=10$ without loss of generality, because the performance of
PAMC is mostly sensitive to the product of $N_{\rm S} N_T$ near optimum.
For example, two PAMC simulations with $\{N_{\rm S}=10, N_T\}$ and
$\{N_{\rm S}=1, 10N_T\}$ are similar in efficiency, if $N_T$ is
reasonably large. The amount of work of a PAMC simulation in terms of
sweeps is $W=R N_{\rm S} N_T$, where $R$ is the average population size.

As shown in Ref.~\cite{wang:15e}, the quality of thermalization of any
thermodynamic observable is in direct correlation with the family
entropy $S_{\rm f}$ and the entropic family size $\rho_{\rm s}$. The
systematic errors, on the other hand, are controlled by the equilibrium
population size $\rho_{\rm f}$. What we here refer to as ``efficiency'' or
``speed-up'' relates to reducing the
statistical as well as the systematic errors while keeping the computational
effort constant. 
Thus, it would be reasonable to use these
quantities as measures of optimality for various PAMC implementations.
$S_{\rm f}$, $\rho_{\rm s}$, and $\rho_{\rm f}$ are defined as
\begin{eqnarray}
S_{\rm f} &=& -\sum_i \nu_i \ln \nu_i, \\
\rho_{\rm s} &=& \lim_{R \rightarrow \infty} R/e^{S_{\rm f}}, \\
\rho_{\rm f} &=& \lim_{R \rightarrow \infty} R \times \rm{var}(\beta F),
\end{eqnarray}
where $\nu_i$ is the fraction of the population that has descended from
replica $i$ in the initial population, and $\beta$ and $F$ are the
inverse temperature and free energy of the system, respectively. The
free energy is measured using the free-energy perturbation method.
Intuitively, $\exp(S_{\rm f})$ characterizes the number of surviving
families and $\rho_{\rm s}$ the average surviving family size.
For a set of simulation parameters, the larger $\rho_{\rm s}$ and $\rho_{\rm
f}$, or the smaller $S_{\rm f}$, the computationally harder the
instance. Keep in mind that $\rho_{\rm f}$ is computationally more expensive to
measure, because many independent runs (at least $10$) are needed to
measure the variance of the free energy.  Note that $S_{\rm f}$ is
``extensive'' and asymptotically grows as $\log(R)$, while both
$\rho_{\rm s}$ and $\rho_{\rm f}$ are ``intensive'' quantities, growing
asymptotically independent of $R$ when $R$ is sufficiently large. 
In our simulations, these metrics are estimated using finite but large-enough $R$ values such that the systematic errors are negligible.

It can be shown \cite{wang:15e,amey:18} that the systematic errors in
any population annealing observable at the limit of large $R$ are
proportional to ${\rm var}(\beta F)$. Therefore, in order to ensure that
the simulations are not affected by the systematic errors, one needs to
make certain that the quantity $\rho_{\rm f}/R$ is sufficiently small.
When well defined, $\rho_{\rm s}$ is strongly correlated with $\rho_{\rm
f}$ \cite{wang:15e} as it is the case for the majority of the spin-glass
instances that we study in this paper. Hence, we may alternatively
minimize $\rho_{\rm s}/R$ or equivalently maximize $S_{\rm f}$ as a
proxy for the quality of equilibration. In our simulations, we ensure
that $S_{\rm f} \gtrsim 2$ for all the instances.

\begin{table}[th!]
\caption{
Simulation parameters for various experiments to optimize PAMC: Spin
selection methods (SSM), annealing schedules (AS), number of
temperatures tuning (NT), dynamic population size experiment (DPS), and
cluster algorithms (CA). $D$ is the space dimension, $L$ is the linear
system size, $R$ is the population size, $T_{\rm{min}}
=1/\beta_{\rm{max}}$ is the lowest temperature simulated, $N_T$ is the
number of temperatures, and $M$ is the number of disorder realizations
studied. The label ``Schedule'' refers to the annealing schedule used,
such as the linear-in-$\beta$ (LB) or the linear-in-$\beta$
linear-in-$T$ (LBLT) schedules. $N_{\rm S}=10$ sweeps are applied to
each replica at each temperature. Note that in the case of dynamic
population sizes (DPS), $R$ is the mean population size. See the text
for more details.  \label{table:simulation-parameters}
}
\begin{tabular*}{\columnwidth}{@{\extracolsep{\fill}} l c c c c c c r}
\hline
\hline
Technique & $D$ & $L$  & $R$  & $T_{\rm{min}}$ & $N_T$ &Schedule & $M$ \\
\hline
SSM & $3$ & $4$  & $5\times10^4$ & $0.2$  & $101$    & LB      & $1000$ \\
SSM & $3$ & $6$  & $2\times10^5$ & $0.2$  & $101$    & LB      & $1000$ \\
SSM & $3$ & $8$  & $5\times10^5$ & $0.2$  & $201$    & LB      & $1000$ \\
SSM & $3$ & $10$ & $1\times10^6$ & $0.2$  & $301$    & LB      & $1000$ \\
AS  & $3$ & $8$  & $5\times10^5$ & $0.2$  & $201$    & All     & $1000$ \\
AS  & $3$ & $10$ & $1\times10^6$ & $0.2$  & $301$    & All     & $1000$ \\
NT  & $2$ & $8$  & $5\times10^4$ & $0.2$  & variable & LBLT    & $100$  \\
NT  & $2$ & $16$ & $2\times10^5$ & $0.2$  & variable & LBLT    & $100$  \\
NT  & $2$ & $25$ & $5\times10^5$ & $0.2$  & variable & LBLT    & $100$  \\
NT  & $2$ & $32$ & $1\times10^6$ & $0.2$  & variable & LBLT    & $100$  \\
NT  & $3$ & $4$  & $5\times10^4$ & $0.2$  & variable & LBLT    & $100$  \\
NT  & $3$ & $6$  & $2\times10^5$ & $0.2$  & variable & LBLT    & $100$  \\
NT  & $3$ & $8$  & $5\times10^5$ & $0.2$  & variable & LBLT    & $100$  \\
NT  & $3$ & $10$ & $1\times10^6$ & $0.2$  & variable & LBLT    & $100$  \\
DPS & $3$ & $6$  & $2\times10^5$ & $0.2$  & $101$    & LB      & $1000$ \\
DPS & $3$ & $8$  & $5\times10^5$ & $0.2$  & $201$    & LB      & $1000$ \\
DPS & $3$ & $10$ & $1\times10^6$ & $0.2$  & $301$    & LB      & $1000$ \\
CA  & $2$ & $8$  & $5\times10^4$ & $0.2$  & $101$    & LB/LBLT & $1000$ \\
CA  & $2$ & $16$ & $2\times10^5$ & $0.2$  & $101$    & LB/LBLT & $1000$ \\
CA  & $2$ & $25$ & $5\times10^5$ & $0.2$  & $201$    & LB/LBLT & $1000$ \\
CA  & $2$ & $32$ & $1\times10^6$ & $0.2$  & $301$    & LB/LBLT & $1000$ \\
CA  & $3$ & $4$  & $5\times10^4$ & $0.2$  & $101$    & LB/LBLT & $1000$ \\
CA  & $3$ & $6$  & $2\times10^5$ & $0.2$  & $101$    & LB/LBLT & $1000$ \\
CA  & $3$ & $8$  & $5\times10^5$ & $0.2$  & $201$    & LB/LBLT & $1000$ \\
CA  & $3$ & $10$ & $1\times10^6$ & $0.2$  & $301$    & LB/LBLT & $1000$ \\
\hline
\hline
\end{tabular*}
\end{table} 

\subsection{Outline of cluster updates used}

Having outlined PAMC, we now briefly introduce the different cluster
algorithms we have experimented with in order to speed up
thermalization.

\subsubsection{Wolff cluster algorithm} 

The Wolff algorithm \cite{wolff:89} greatly speeds up simulations of
Ising systems without frustration near the critical point.  It is well
known that the Wolff algorithm does not work well for spin glasses in 3D
\cite{kessler:90} because the cluster size grows too quickly with
$\beta$. Nevertheless, we revisit this algorithm systematically in both
2D and 3D. The idea is that even if the cluster size grows too quickly
when $\beta$ is still relatively small, the mean cluster size
(normalized by the number of spins $N_{\rm spins}$) is still a
continuous function in the range $[0,1]$ when $\beta$ grows from
$\beta=0$ to $\infty$.  Therefore, it is a reasonable question to ask if
there would be some speed-up when restricting the algorithm to the
temperature range where the normalized mean cluster size is neither too
larger nor too small, for example, in the range $[0.1,0.9]$.

In the ferromagnetic Ising model, where $J_{ij}=J=1$, one adds a
neighboring spin $S_j$ when it is parallel to a spin $S_i$ in the
cluster with probability $p_{\rm{c}}=1-\exp(-2J\beta)$. In spin glasses,
this is generalized as follows: One adds a neighboring spin $S_j$ to
$S_i$ when the bond between the two spins is satisfied and with
probability $p_{\rm{c}}=1-\exp(-2|J_{ij}|\beta)$. This can be compactly
written as $p_{\rm{c}}=\max[0,1-\exp(-2\beta J_{ij}S_iS_j)]$
\cite{kessler:90}. Note that from $p_{\rm{c}}$, there are two
interesting limits for the mean cluster size. In the limit $\beta
\rightarrow 0$, the average cluster size is clearly $0$, and in the
limit $\beta \rightarrow \infty$, the normalized cluster size tends to
$1$, because in the ground state each spin has at least one satisfied
bond with its neighbors and all the spins would be added to the cluster.
From the expression for $p_{\rm{c}}$ one can see that frustration
actually makes the cluster size grow \textit{slower} as a function of
$\beta$.  However, frustration also significantly reduces the transition
temperature, which is the primary reason why the Wolff algorithm is less
efficient for spin glasses. Finally, note that the Wolff algorithm is
both ergodic and satisfies detailed balance.

\subsubsection{Houdayer cluster algorithm} 

Designed for spin glasses, the Houdayer cluster algorithm
\cite{houdayer:01} or its generalization, the isoenergetic cluster moves
(ICM) \cite{zhu:15}, greatly improves the sampling for parallel tempering
in 2D, while less so in 3D.  ICM in 3D, like the Wolff algorithm, is
restricted to a temperature window where the method is most efficient
\cite{zhu:15}.  ICM works by updating two replicas at the same time.
First, an overlap between the two replicas is constructed, which
naturally forms positive and negative islands. One island is selected,
and the spin configurations of the island in both replicas are flipped.

In its original implementation, the spin down sector is always used to
construct the cluster. In the implementation of Zhu {\em et al.}~a full
replica is flipped if the chosen island is in the positive sector to
make it negative \cite{zhu:15} and therefore reduce the size of the
clusters.  Here, we improve on this implementation by allowing the
chosen island to be either positive or negative, and flipping the spins
of the island in both replicas. Therefore, we never flip a full replica.
This saves computational time and also has the advantage that it does
not artificially make the spin overlap function symmetric. ICM satisfies
detailed balance but is not ergodic.  Therefore, the algorithm is
usually combined with an ergodic method such as the Metropolis
algorithm. ICM greatly improves the thermalization time, and also
slightly improves the autocorrelation time in parallel tempering.
Because PAMC is a sequential method, there is no thermalization stage.
We therefore focus on whether the algorithm reduces correlations, i.e.,
systematic and statistical errors. Our implementation of PAMC with ICM
is as follows: First, after each resampling step, we do regular Monte
Carlo sweeps and ICM updates alternately. We first do $N_{\rm S}/2$
lattice sweeps for each replica, followed by $R$ ICM updates done by
randomly pairing two replicas in the population, followed by another
$N_{\rm S}/2$ lattice sweeps.  Second, for each ICM update, we choose an
island from the spin sector with the smaller number of spins. Then the
spin configurations of the island in both replicas are flipped. This
effectively means that the spin configurations associated with the
selected island are either exchanged or flipped depending on the sign of
the island being negative in the former or positive in the latter. Note
that the combined energy of the two replicas is conserved in both cases,
therefore making the algorithm rejection free.

\begin{figure}[t!]
\begin{center}
\includegraphics[width=\columnwidth]{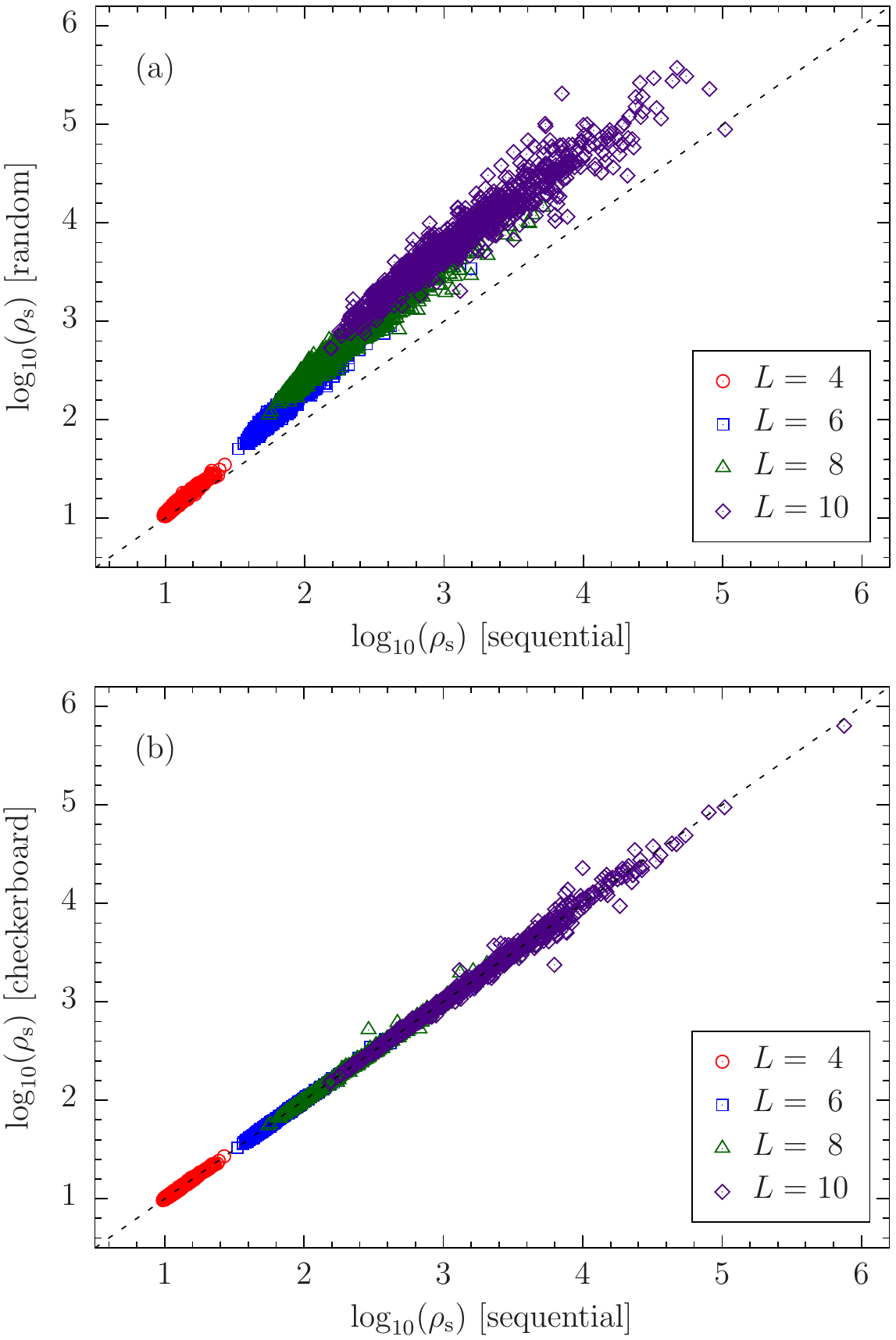}
\caption{
Comparison of the entropic population size $\rho_{\rm s}$ for different
spin selection methods: random, sequential and checkerboard updates in
three space dimensions.  Sequential and checkerboard updates have
similar efficiency (b), and both are more efficient than random
updates (a).
}
\label{SSM}
\end{center}
\end{figure}

\section{Implementation optimizations}
\label{imp}

In this section, we present our implementation improvement to the
population annealing algorithm. We first present spin selection methods,
followed by experiments using different annealing schedules, numbers of
temperatures, and the use of a dynamic population.  The simulation
parameters are summarized in Table~\ref{table:simulation-parameters}.

\subsection{Comparison of spin selection methods}

We have studied three spin selection methods: sequential, random, and
checkerboard. We have carried out a large-scale simulation in 3D to
compare these methods for $L=4$, $6$, $8$, and $10$, with $1000$
instances for each system size. We first run the simulations using the
parameters in Table~\ref{table:simulation-parameters}. To measure
$S_{\rm f}$ or $\rho_{\rm s}$ reliably, we require $S_{\rm f} \gtrsim 2$
\cite{wang:15e}. When this is not satisfied for a particular instance,
we rerun it with a larger population size. We then compare $\rho_{\rm
s}$ at the lowest temperature between different spin selection methods.
Figure \ref{SSM} shows scatter plots comparing $\rho_{\rm s}$ instance
by instance for different system sizes and using different spin
selections methods. Figure \ref{SSM}(a) compares random to sequential
updates, whereas Fig.~\ref{SSM}(b) compares checkerboard to sequential
updates.  Interestingly, sequential and checkerboard updates have
similar efficiency (the data lie on the diagonal), whereas both
sequential and checkerboard are more efficient than random updates.
This is particularly visible for the larger system sizes, e.g., $L =
10$. The random selection method is therefore the least efficient update
technique for disordered Boolean problems, keeping in mind that it
requires the computation of an additional random number for each
attempted spin update thus slowing down the simulation. We surmise that
a sequential updating of the spins accelerates the mobility of domain
walls in most cases.  However, in some pathological examples, such as the
one-dimensional Ising chain random updating is needed for Monte Carlo to
be ergodic.

\begin{figure*}[t!]
\begin{center}
\includegraphics[width=7in]{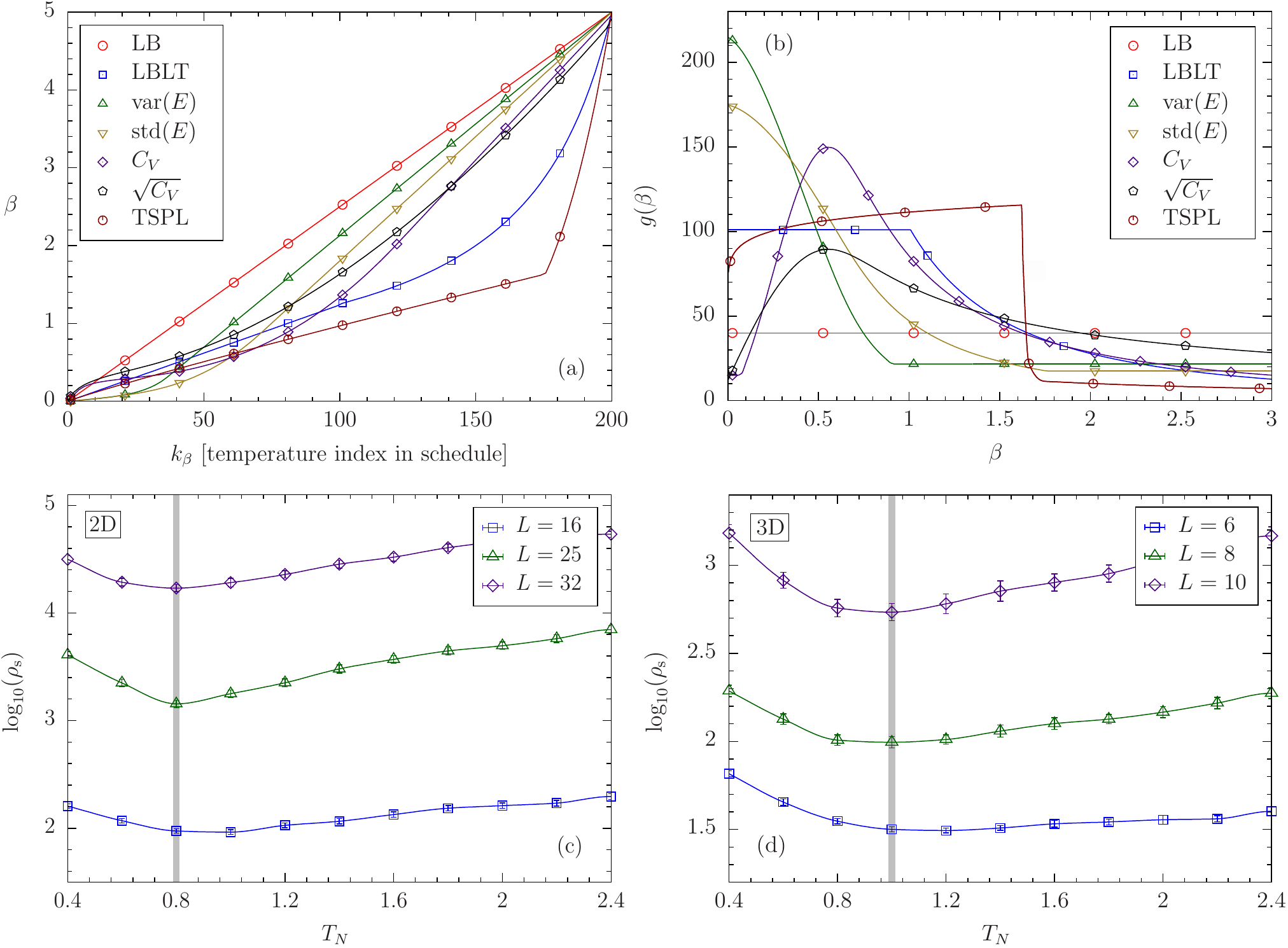}
\caption{
Panel (a) shows the $\beta$ values as a function of the inverse
temperature index $k_\beta$ for the different schedules experimented
with and panel (b) shows the resulting $\beta$ densities $g(\beta)$
(the data are cut off at $\beta = 3$ for clarity).  Note that both TSPL
and LBLT schedules have more temperatures at high $T$.  Panels (c) and
(d) show $\rho_{\rm s}$ as a function of $T_N$ for 2D and
3D simulations, respectively. The vertical shaded line
marks the optimum.  See the main text for details.
}
\label{schedule}
\end{center}
\end{figure*}

\subsection{Optimizing annealing schedules}

Most early population annealing simulations used a simple
linear-in-$\beta$ (LB) schedule where the change in $\beta$ in the
annealing schedule is constant as a function of the temperature index.
This, however, is not necessarily the most optimal schedule to use.  We
use two approaches to optimize the annealing schedules and the number of
temperatures: One approach uses a mathematical model with free
parameters to be optimized and the other includes adaptive schedules
based on a guiding function, e.g., the energy fluctuations or the
specific heat.  For the parametric schedules we introduce a
linear-in-$\beta$ linear-in-$T$ (LBLT) and a two-stage power-law
schedule (TSPL).  For the LBLT schedule there is one parameter to tune,
namely a tuning temperature $T_N$ \cite{comment:tmin}.  In this
schedule, half of the temperatures above $T_N$ are linear in $\beta$,
while the other half below $T_N$ are linear in $T$.  For the TSPL
schedule we define a rescaled annealing time $\tau=k_\beta/(N_T-1) \in
[0,1]$, where $k_\beta$ is the annealing step (or temperature index)
$0$, \ldots, $N_T-1$. The TSPL schedule is modeled as
\begin{eqnarray}
\beta(\tau)=
a \tau^{\alpha_1}\theta(\tau_0-\tau)+b \tau^{\alpha_2}\theta(\tau-\tau_0),
\end{eqnarray}
where $\theta$ is the Heaviside step function. Here $\alpha_1$ and
$\alpha_2$ are free parameters. $a$ and $b$ enforce continuity and the
final annealing temperature. $\tau_0$ is selected to enforce a
switch-over temperature $\beta_0$.  We optimize the LBLT schedule with a
simple scan of the parameter $T_N$.  The optimum value of $T_N$ (where
$\rho_{\rm s}$ is minimal) is shown in Fig.~\ref{schedule}(c) for 2D
($T_N \approx 0.8$, marked with a vertical shaded area) and
Fig.~\ref{schedule}(d) for 3D ($T_N \approx 1.0$, marked with a vertical
shaded area). 

\begin{figure}[ht!]
\begin{center}
\includegraphics[width=\columnwidth]{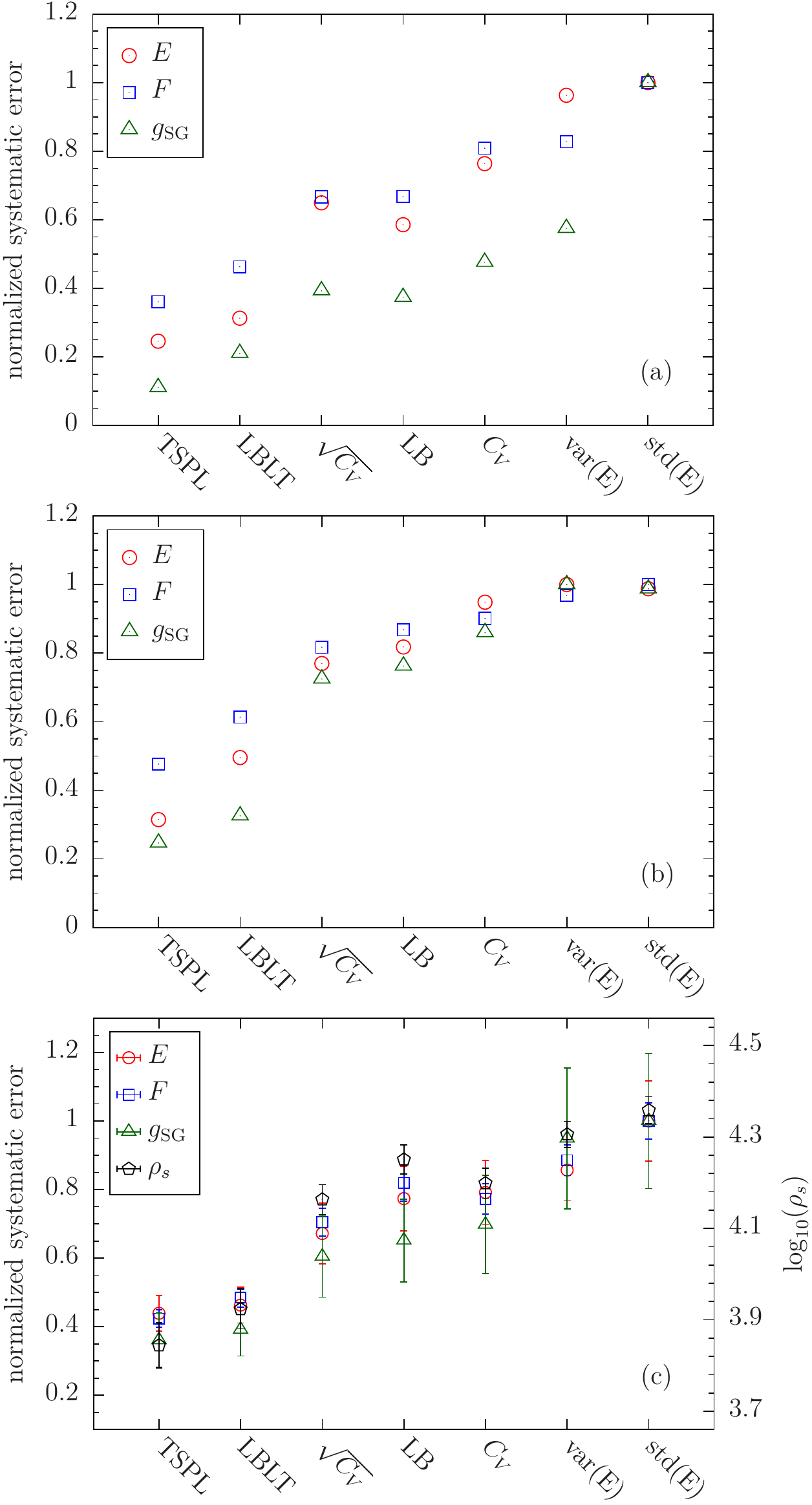}
\caption{
Comparison of the systematic errors for various annealing schedules. The
studied observables are energy ($E$), free energy ($F$), and the spin
glass Binder cumulant ($g_{\rm SG}$) for the system size $L=10$. Panels
(a) and (b) show the systemic errors for two randomly chosen hard
instances, whereas panel (c) illustrates the systematic errors averaged
over 100 of the hardest instances. Systematic errors of different
observable often have magnitudes largely apart. For this reason, the
errors in each observable have been normalized relative to the maximum
error across all schedules. For instance, in the top panel the ${\rm
std}(E)$ schedule which has the greatest systematic error is normalized
to 1 while the rest of the schedules lie below 1. It is seen from the
plots that the TSPL schedule is the most efficient. The LBLT schedule,
although conveniently simple, competes well with the optimal schedule.
Note that we also show $\rho_s$ (as a dual $y$-axis) in panel (c). We
observe that $\rho_s$ greatly correlates with the systematic errors
justifying the use of it as an effective optimization criterion.}
\label{schedule-performance}
\end{center}
\end{figure}

\begin{figure}[ht!]
\begin{center}
\includegraphics[width=\columnwidth]{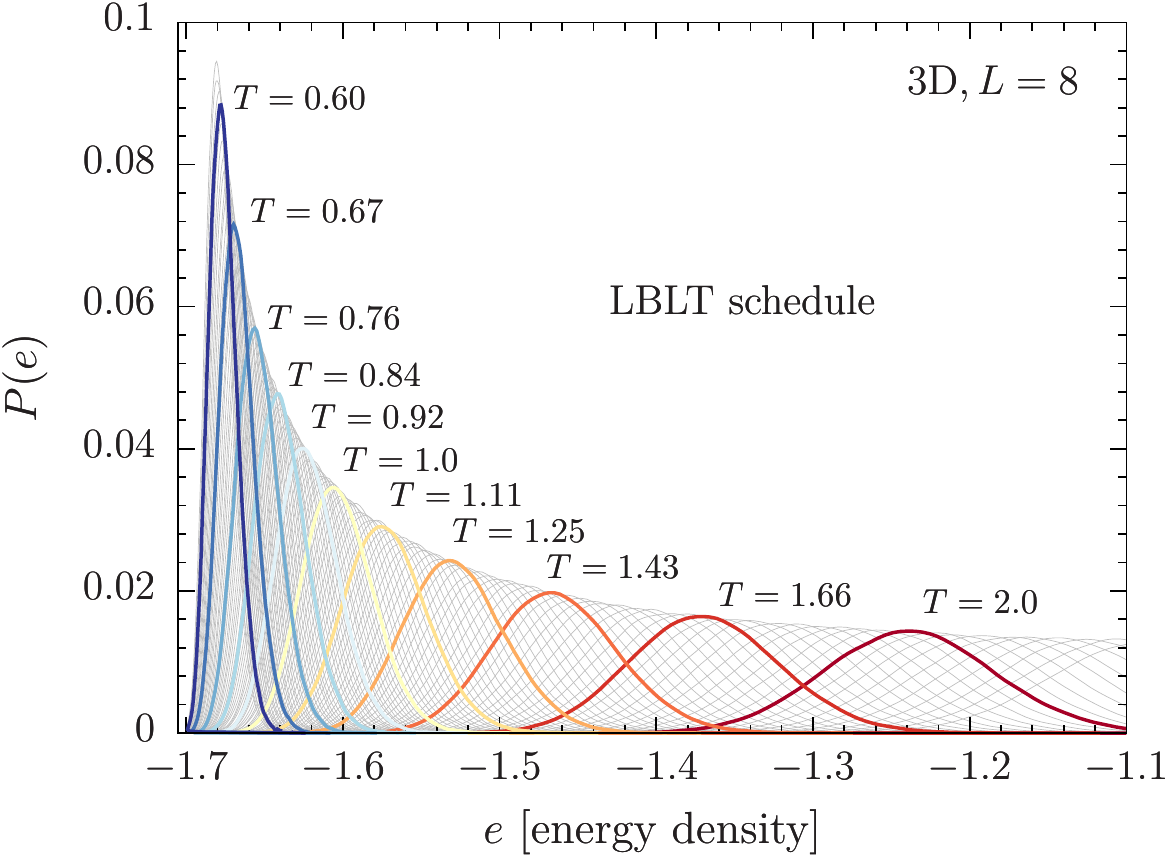}
\caption{
Energy density distribution of the LBLT annealing schedule for $L=8$ in
three space dimensions. Thinner curves show the histograms at all
temperatures whereas the thicker ones are drawn at every $10$
temperature steps. There are $200$ temperature steps in total. The
histograms overlap considerably.
}
\label{energy_overlap}
\end{center}
\end{figure}

The TSPL schedule, however, has more parameters that have to be tuned.
Therefore, we have used the Bayesian optimization package Spearmint
\cite{snoek:12,adams:16} rather than a full grid scan in the entire
parameter space. We find numerically that the parameters $\alpha_1 =
\exp(-0.0734)$, $\alpha_2 = \exp(2.15)$, and $\beta_0=1.63$ work well.
However, we note that there is no guarantee of global optimality. 
For the adaptive schedules, we optimize using information provided by
energy fluctuations, because energy is directly related to the
resampling of the population. We therefore define a density of inverse
temperature $\beta$, $g(\beta)$, and study the following adaptive
schemes.
\begin{enumerate}
\item[] var($E$) schedule with $g(\beta) \sim {\rm{var}}(E)$,
\item[] std($E$) schedule with $g(\beta) \sim \sqrt{\rm{var}(E)}$,
\item[] $C_V$ schedule with $g(\beta) \sim C_V(\beta)$,
\item[] $\sqrt{C_V}$ schedule with $g(\beta) \sim \sqrt{C_V(\beta)}$,
\end{enumerate}
where $C_V$ is the specific heat of the system. Note that the functions
are disorder averaged, and the proportionality is determined by the
number of temperatures. Because $g(\beta)$ may become extremely small,
we have replaced all the function values that are less than $10\%$ of
$\max(g)$ by $0.1 \times \max(g)$ to prevent large temperature leaps.
With this small modification, we generate $N_T$ temperatures according
to the above density functions. The shapes and $\beta$ densities of all
schedules are shown in Figs.~\ref{schedule}(a) and \ref{schedule}(b),
respectively. There are clear differences between the different
schedules, especially in comparison to the traditionally used LB
schedule.  We compare the efficiency of these different schedules in
Fig.~\ref{schedule-performance} by analyzing the systematic errors in a
number of paradigmatic observables. We have studied the internal energy
($E$), free energy ($F$), and the spin-glass Binder cumulant
\cite{binder:81} for the system size $L=10$. To overcome the scale
difference when showing the systematic errors for these observables in
one plot, we have normalized the errors with respect to the schedule
that has the greatest error.  Therefore, all the errors will be relative
to that of the worst schedule.  In Figs.~\ref{schedule-performance}(a)
and \ref{schedule-performance}(b) we show the normalized systematic
errors for two randomly chosen and extremely hard instances. In
Fig.~\ref{schedule-performance}(c) we show the disorder averaged
systematic errors calculated from $100$ of the hardest instances. It can
be readily seen from the plots that the LBLT and TSPL schedules yield
the best efficiencies among all the experimented schedules with TSPL
slightly more efficient.  Both LBLT and TSPL schedules place more
temperatures at high temperature values (smaller $\beta$ values),
presumably because the Metropolis dynamics is more effective at high
temperatures. Additionally, in Fig.~\ref{schedule-performance}(c) we
have shown $\rho_s$ for various schedules. We observe great correlation
between $\rho_s$ and the systematic errors which corroborates the use of
$\rho_s$ as a good measure of efficiency.

\begin{figure}[t!]
\begin{center}
\includegraphics[width=\columnwidth]{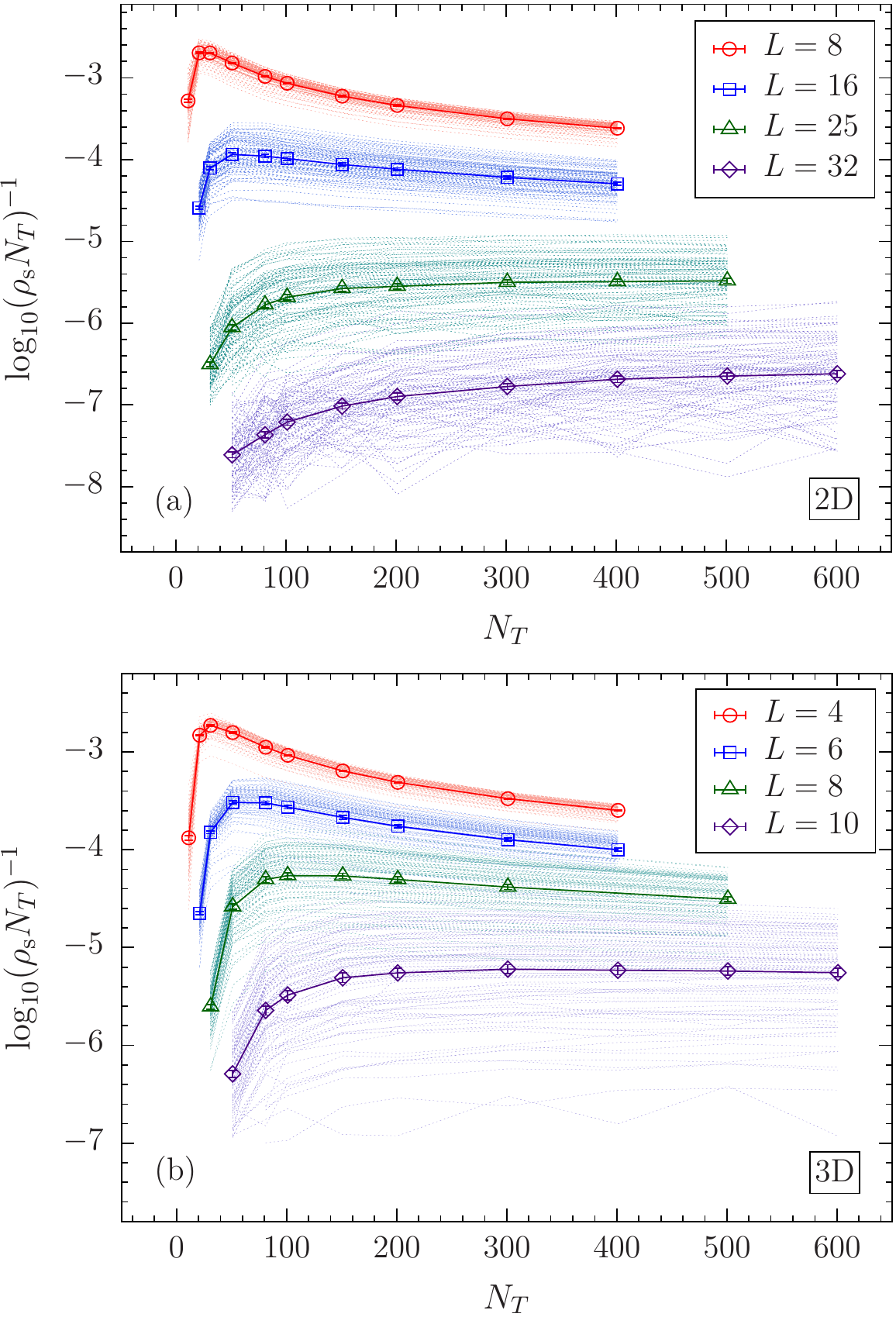}
\caption{
Optimization of the number of annealing steps $N_T$ in two space
dimensions [2D, panel (a)] and three space dimensions [3D, panel (b)].
To maximize sampling efficiency, one needs to optimize $1/(\rho_{\rm
s}N_T)$ with respect to $N_T$.  In both panels the points and the solid
curves show the disorder average while the dashed envelopes display all
$100$ studied instances. For smaller system sizes the peak (optimum) is
sharp, whereas for systems with more than approximately $1000$ spins the
peak is broadened, especially in two dimensions. The reason for this
broadening can be understood by noticing the increase in the density of
chaotic samples as the system grows in size (wiggly lines).
}
\label{NT}
\end{center}
\end{figure}

We stress that the optimum schedule depends on the choice of the number
of sweeps at each anneal step $N_{\rm S}$, because $N_T$ and $N_{\rm S}$
are exchangeable when $N_T$ is large enough. In our approach, we have
fixed $N_{\rm S}$.  It is therefore possible that other techniques may
result in different optimal schedules. For instance, one may use the
energy distribution overlaps at two temperatures to define the optimum
schedule \cite{amey:18,barash:17a}, which only depends on the
thermodynamic properties of the system. As an example, in Fig.
\ref{energy_overlap} we show the energy distributions of the LBLT
schedule for $L=8$ in 3D. The energy histograms overlap considerably up
to several temperature steps.  Within this framework, the optimization
is transferred to the distribution of sweeps. However, the density of
work (the product of density of $\beta$ and density of sweeps) should be
similar in the two different approaches. In our implementation as the
number of sweeps is constant, the density of work is the same as the
density of $\beta$.

\subsection{Optimization of the number of temperatures}

To optimize the number of temperatures and their range, we use the LBLT
schedule as it is easy to implement and very close to optimal.  Our figure of
merit is to maximize the number of independent measurements $R/\rho_{\rm
s}$ for constant work $W=R N_{\rm S} N_T$.  We define efficiency as
$\gamma = R/(\rho_{\rm s} W)$ by tuning $N_T$ for a constant $W$.
Because $N_{\rm S}=10$ is fixed, we need to maximize $1/(\rho_{\rm
s}N_T)$ by tuning $N_T$. In the limit $R \rightarrow \infty$, $\rho_{\rm
s}$ and the efficiency $\gamma$ are independent of the population size.
This is expected as $\gamma$ is an intensive quantity. Therefore, to
measure $\gamma$, we only need to make sure $R$ is sufficiently large
such that $\rho_{\rm s}$ has converged.  It is not necessary to use the
same $W$ for different $N_T$.

The results for both two- and three-dimensional systems are shown in
Figs.~\ref{NT}(a) and \ref{NT}(b), respectively.  The solid curves show
the disorder average while the dashed envelopes are the
instance-by-instance results. It is interesting to note that for
relatively smaller system sizes we observe a pronounced peak. The
existence of an optimum number of temperatures can be intuitively
understood in the following way: For a fixed amount of computational
effort, if $N_T$ is too small, then the annealing or resampling would
become too stochastic, which is inefficient.  On the other hand, if the
annealing is too slow ($N_T$ is too large) this becomes unnecessary and
keeping a larger population size is more efficient. Therefore, the
optimum comes from a careful balance between $N_T$ and $R$. However as
the system size grows, the optimum peak starts to flatten out due to the
onset of temperature chaos
\cite{neynifle:97,aspelmeier:02a,jonsson:02a,katzgraber:07,fernandez:13,wang:15a,zhu:16}.
This can be seen in Fig.~\ref{NT} as a discernible increase in the
density of instances with irregular oscillatory behavior. Thus we
conclude that the optimization presented here, although capturing the
bulk of the instances, might not be reliable in case of extremely hard
(chaotic) instances. Instead one may consider performing more Metropolis
sweeps rather than merely increasing the temperature steps or the
population size. This is especially relevant if memory (which correlates
to $R$) becomes a concern for the hardest instances.

\begin{figure*}[t!]
\begin{center}
\includegraphics[width=\textwidth]{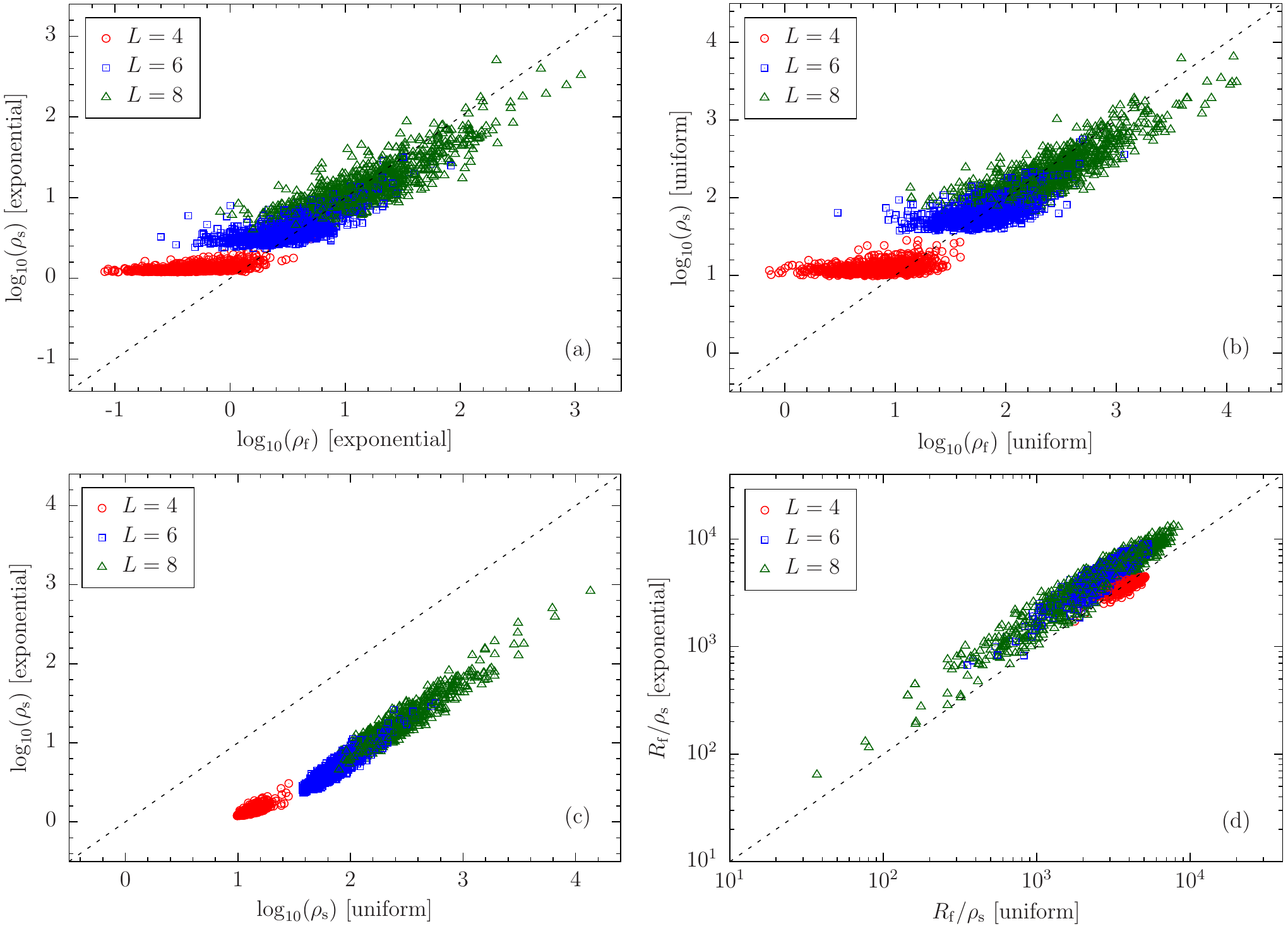}
\caption{
Instance-by-instance comparison for a PAMC simulation with fixed and
dynamic population sizes.  With a dynamic population size, $\rho_{\rm s}$
and $\rho_{\rm f}$ are well correlated, similarly to the case of uniform
population. $\rho_{\rm s}$ is greatly reduced, suggesting that the
simulation is much better at the level of averaging over all
temperatures. The dynamic population size is also more efficient than
the uniform one using the worst-case measure. Here $R_{\rm f}$ is the 
final population size.
}
\label{DPS}
\end{center}
\end{figure*}

\subsection{Dynamic population sizes}

The reason the LBLT schedule is more efficient than a simple LB schedule
is because the Metropolis dynamics is less effective at low
temperatures, and therefore using more ``hotter'' temperatures is more
efficient. Here we investigate another technique, namely a variable number of
replicas that depends on the annealing temperature, thus having a
similar effect to having more temperatures at higher values.  Regular
PAMC is designed to have an approximately uniform population size as a
function of temperature. Here we allow the population size to change
with $\beta$. Because most families are removed at a relatively early
stage of the anneal, transferring some replicas from low temperatures to
high temperatures may increase the diversity of the final population,
even though the final population size would be smaller
\cite{comment:pop}.

We study a simple clipped exponential population schedule where the
population starts as a constant $R_0$ until $\beta=\beta_0$, and then
decreases exponentially to $R_{\rm f}=R_0/r$ at $\beta=\beta_{\rm{max}}$,
\begin{equation}
R(\beta) = 
\begin{cases}
R_0 & \!\!\!\beta \leq \beta_0 \\
a R_0 /\left[(r - 1)(e^{\beta S} - e^{\beta_0 S}) + a\right] & \!\!\!\beta > \beta_0,
\end{cases}
\end{equation}
where $a = \exp(\beta_{\rm{max}}S) - \exp(\beta_0S)$. The free
parameters to tune are $S$, $\beta_0$, and $r$. $S$ is chosen such that
the function is continuous and naturally characterizes the slope of the
curve. Once the parameters are optimized, we can scale the full function
to have a comparable average population size to that of the uniform
schedule. The optimization is again done using Bayesian statistics, and
we obtain $\beta_0=0.9$, $r=33.8$, and $S=\exp(-2.52)$.

\begin{figure*}[th!]
\begin{center}
\includegraphics[width=\textwidth]{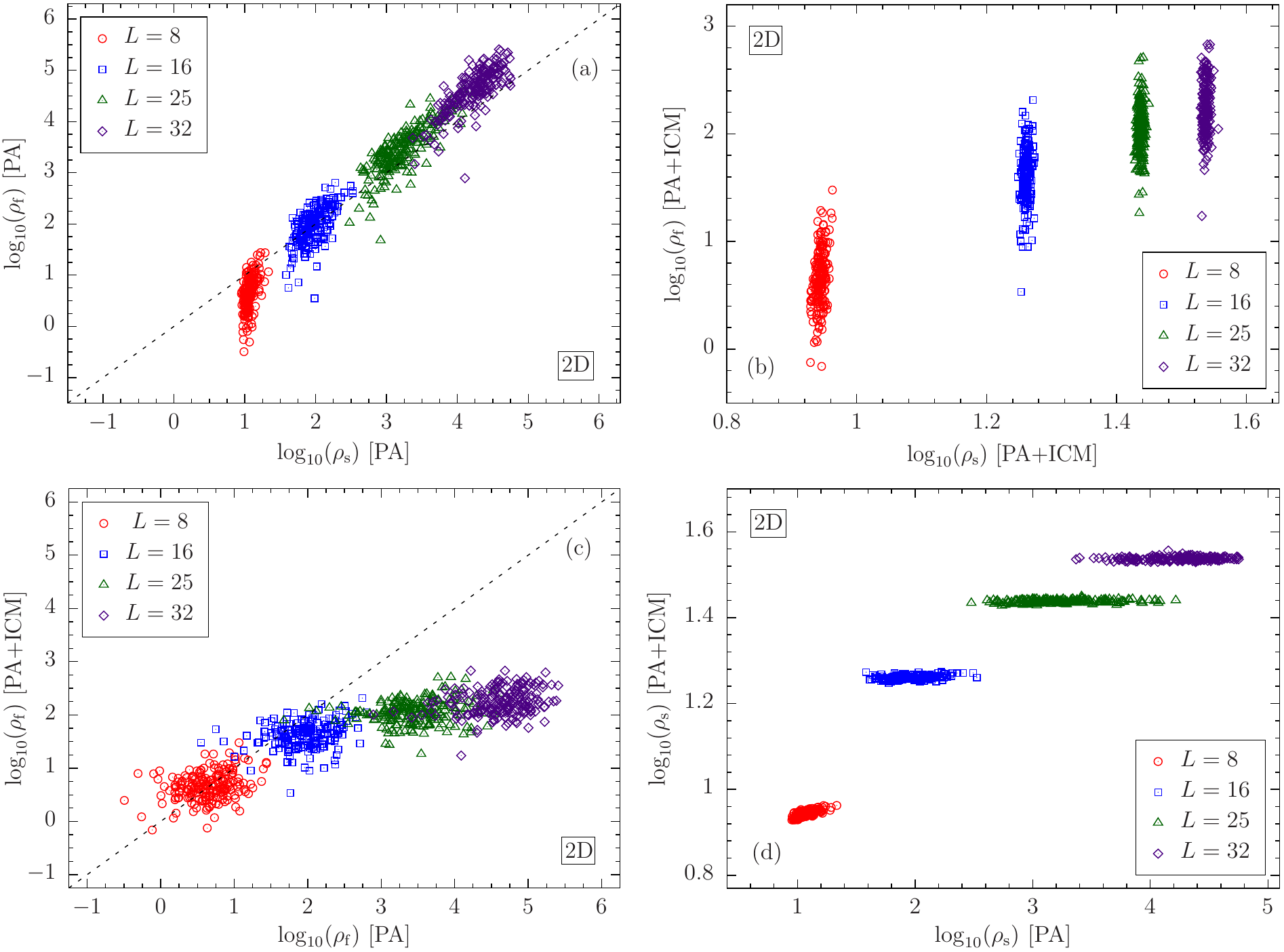}
\caption{
Population annealing with ICM updates in 2D. Note that replica family is not
well defined when ICM updates are included. Therefore, we use $\rho_{\rm f}$
to characterize speed-up. Significant speed-up is observed in 2D. 
}
\label{ICM-2D}
\end{center}
\end{figure*}

It is noteworthy to mention that there are two different measures to
detect efficiency when the population size is allowed to change. For the
same average population size, the dynamic population schedule is
\textit{always} better at high temperature. However, at low temperature,
a smaller $\rho_{\rm s}$ does not justify that the number of independent
measurements is larger, because $R$ is also smaller.  It is thus
reasonable to optimize the parameters using $\rho_{\rm s}$, and then
also to compare to $R/\rho_{\rm s}$. Note that we use the local
population size $R$ at each temperature to compute $\rho_{\rm s}$.  The
correlations and comparisons of $\rho_{\rm s}$ and $\rho_{\rm f}$ are
also studied. With the optimum parameters, we compare the efficiency of
the dynamic and uniform population sizes. The results are shown in
Fig.~\ref{DPS}. We see that $\rho_{\rm s}$ and $\rho_{\rm f}$ are well
correlated for the dynamic population size. $\rho_{\rm s}$ is greatly
reduced, suggesting that the simulation is much better at the level of
averaging over all temperatures. We also see that even using the
worst-case measure, the dynamic population size is more efficient than
the uniform one. Note, however, that the peak memory use of the dynamic
population size is larger due to the nonuniformity of the number of
replicas as a function of $\beta$.

\section{Algorithmic accelerators}
\label{acc}

We now turn our attention to algorithmic accelerators by including
cluster updates in the simulation. The simulation parameters are
summarized in Table~\ref{table:simulation-parameters}.

\begin{figure*}[t!]
\begin{center}
\includegraphics[width=\textwidth]{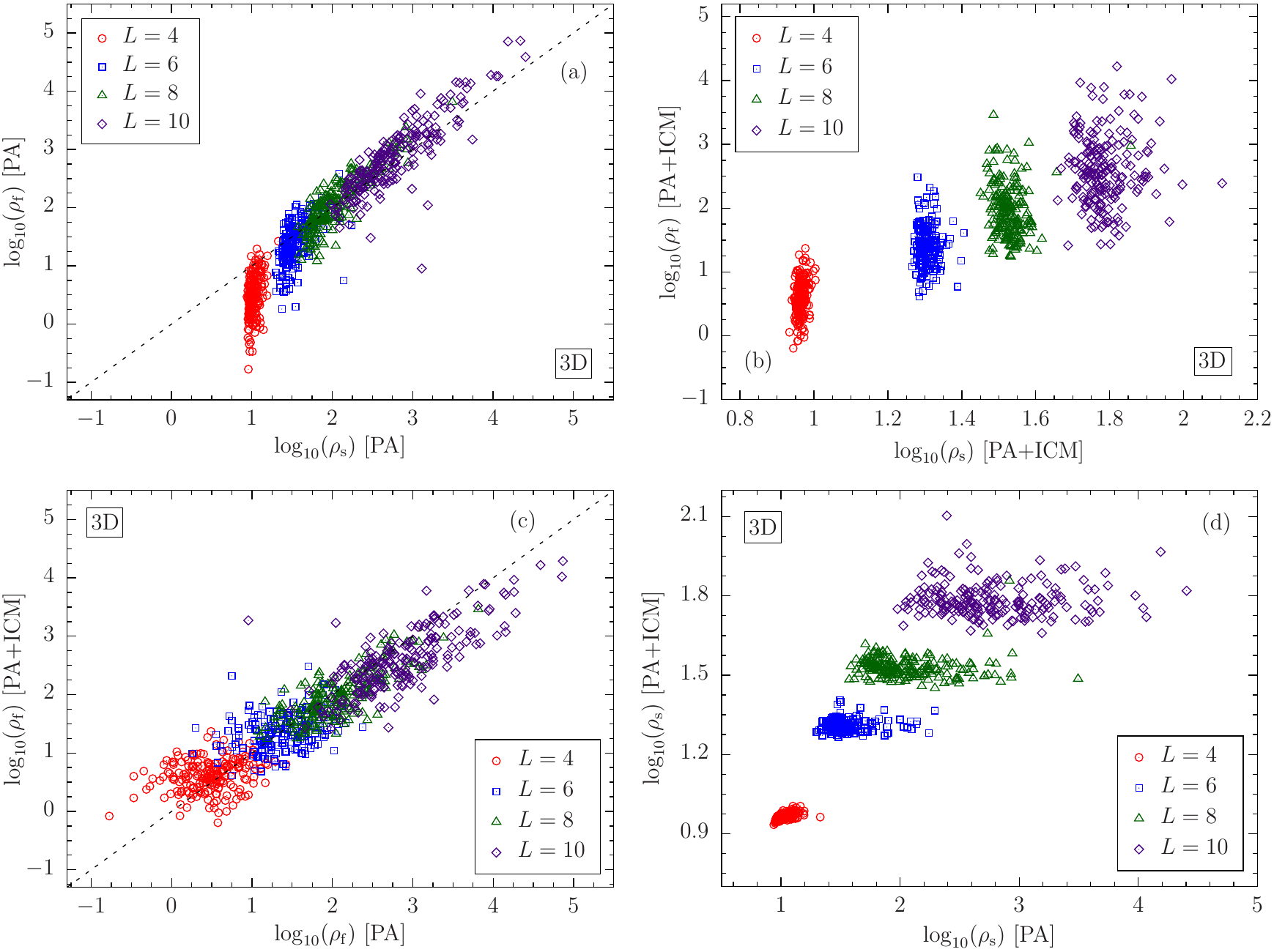}
\caption{
Population annealing with ICM updates in 3D. Note that replica family is not
well defined when ICM updates are included. Therefore, we use $\rho_{\rm f}$
to characterize speed-up. Modest speed-up is observed in 3D.
}
\label{ICM-3D}
\end{center}
\end{figure*}

\subsection{Isoenergetic cluster updates}

Here we study PAMC with ICM updates. In 3D, similarly to the Wolff
algorithm, there is an effective temperature range where ICM (see
Ref.~\cite{zhu:15} for more details) is efficient. In ICM, two replicas
are updated simultaneously. This process uses the detailed structure of
the two replica configurations, and it is natural to question if the
family of a replica is still well defined. For example, occasionally,
two replicas may merely exchange their configurations. This is
equivalent to exchanging their family names which potentially increases
the diversity of the population {\em at little cost}. To resolve and
investigate this issue, we have therefore measured the (computationally
more expensive) equilibration population size $\rho_{\rm f}$ as well,
which unlike $\rho_{\rm s}$, does not depend on the definition of the
families. Our results are shown in Fig.~\ref{ICM-2D} and
Fig.~\ref{ICM-3D}, for 2D and 3D, respectively.  We find that $\rho_{\rm
s}$ is indeed artificially reduced by the cluster updates. In both 2D
and 3D, $\rho_{\rm f}$ has a wide distribution, while $\rho_{\rm s}$ is
almost identical for all instances. Furthermore, $\rho_{\rm s}$ and
$\rho_{\rm f}$ are strongly correlated for regular PAMC, but the
correlation is poor when ICM is turned on. Therefore, we conclude that
$\rho_{\rm s}$ is no longer a good equilibration metric for PAMC when
combined with ICM. Using $\rho_{\rm f}$, we find that similar to PT
\cite{zhu:15}, there is clear speed-up in 2D. In 3D, however, the
speed-up becomes again marginal. This is in contrast to the discernible
speed-up for PT with the inclusion of ICM in 3D. The results suggest
that ICM is mostly efficient in 2D and likely quasi-2D lattices,
reducing both thermalization times (PT) and correlations (PAMC and PT).
In 3D, ICM merely reduces thermalization times, while marginally
influencing correlations.

\subsection{Wolff cluster updates}

Wolff cluster updates are not effective in spin-glass simulations.  We,
nevertheless, have revisited this type of cluster update in the context
of PAMC for the sake of completeness. More details can be found in
Appendix \ref{appendix:Wolff}.

\section{Parallel Implementation} 
\label{parallel}

Population annealing is especially well suited for parallel computing
because operations on the replicas can be carried out independently and
communication is minimal. Since OpenMP is a shared-memory
parallelization library, it is limited to the resources available on a
{\em single node} of a high-performance computing system. Although
modern compute nodes have many cores and large amounts of RAM, these are
considerably smaller than the number of available nodes by often several
orders of magnitude. To benefit from machines with multiple compute
nodes and therefore simulate larger problem sizes, we now present an MPI
implementation of PAMC which can utilize resources up to the size of the
cluster.  While for typical problem sizes single-node OpenMP
implementations might suffice for the bulk of the studied instances,
hard-to-thermalize instances could then be simulated using a massively
parallel MPI implementation with extremely large population sizes.
Although the exact run time depends on many variables such as the
simulation parameters, architecture, code optimality, compiler, etc.,
here we show some example of a typical simulation time with the
parameters listed in Table \ref{table:simulation-parameters}.  On a
20-core node with Intel Xeon E5-2670 v2 2.50 GHz processors, it takes
approximately $1.3$, $12$, and $75$ minutes to simulate an instance in
3D with $N=216$, $512$, and $1000$ spins, respectively.

\subsection{Massively parallel MPI implementation}

\begin{figure}[t!]
\begin{center}
\includegraphics[width=\columnwidth]{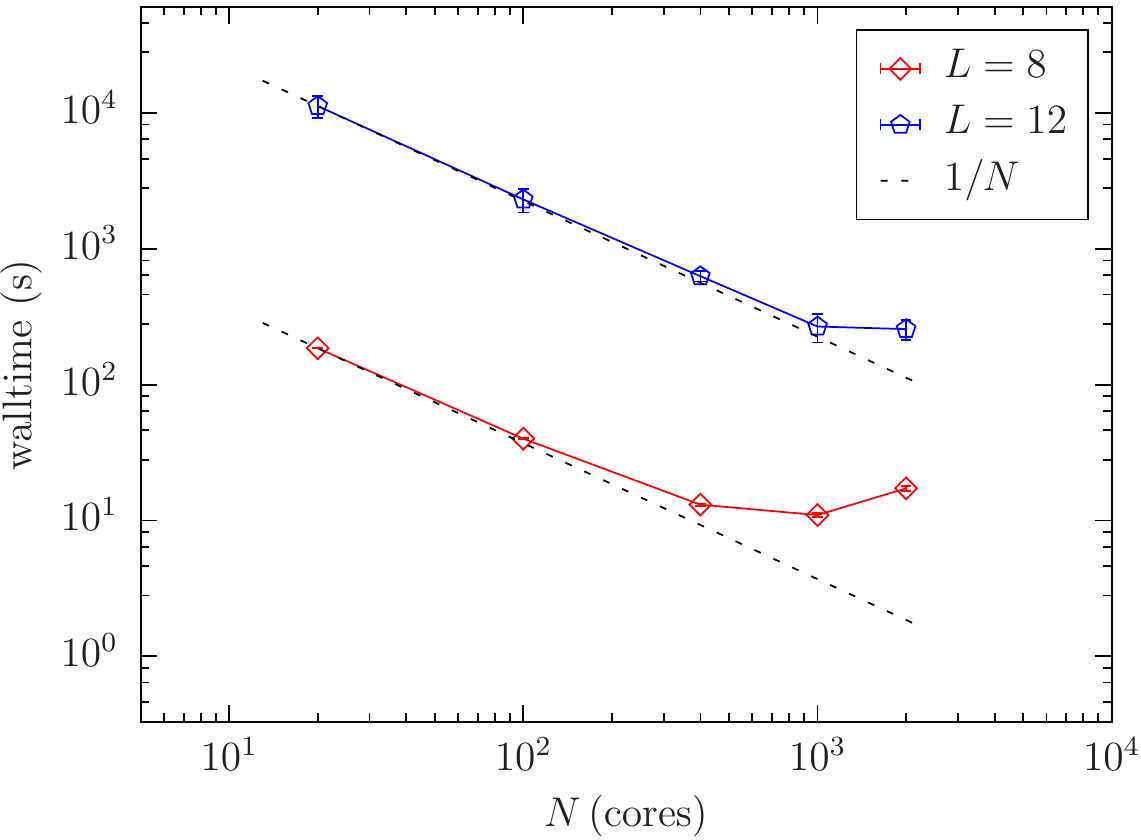}
\caption{
Scaling of the total wall time as a function of the number of processors
$N$ for two system sizes $L=8$ and $L=12$. Launching and initialization
time are not included. Note that the efficiency becomes better for larger
and harder problems. For $L=12$, the scaling remains $1/N$ up to about
1000 processors. The efficiency then decreases when the time for
collecting observables becomes dominant. Note that resampling still
takes a relatively small time.
}
\label{Benchmark}
\end{center}
\end{figure}

The performance and scaling of our MPI implementation for 3D
Edwards-Anderson spin glasses is shown in Fig.~\ref{Benchmark}. Note
that the wall time scales $\sim 1/N$ with $N$ the number of cores for
less than $1000$ cores.  In our implementation, the population is
partitioned equally between MPI processes (ranks). Each rank is assigned
an index $k$ with I/O operations occurring on the 0th rank. A rank has a
local population on which the Monte Carlo sweeps and resampling are
carried out. We also define a global index $G$ which is the index of a
replica as if it were in a single continuous array. In practice, the
global index $G$ of a replica $j$ on a rank $k$ is computed as the sum of the local
populations $r_i$ on the preceding ranks plus the local index $j$, i.e., 
\begin{eqnarray}
G = j + \sum\limits_{i=0}^{k-1}r_i.
\end{eqnarray}
The global index for a particular replica varies as its position in the
global population changes.

Load balancing is carried out when a threshold percentage between the
minimum and maximum local populations is exceeded. In our
implementation, all members of a family must be in a continuous range of
global indices to allow for efficient computation of the family entropy
and the overlap function of the replicas.  Therefore, load balancing
must maintain adjacency. The destination rank $k$ of a replica is
determined by evenly partitioning the global population such that each
rank has approximately the same number of replicas, i.e.,
\begin{eqnarray}
k = \lfloor G / \left(\tfrac{R}{N}\right) \rfloor ,
\end{eqnarray}
where $N$ is the number of ranks (cores).

Measurement of most observables is typically an efficient accumulation
operation, i.e.,
\begin{eqnarray}
\langle \mathcal{A} \rangle = \frac{1}{R}\sum\limits_{k}^{N} \sum\limits_{j}^{r_k}\mathcal{A}_{j,k}.
\end{eqnarray}
On the other hand, measuring observables such as the spin-glass
overlap is more difficult and only done at select temperatures. Sets of
replicas are randomly sampled from a rank's local population and copies
are sent to the range of ranks $[(k+N/4) \mod N, (k+3N/4) \mod N]$ with
periodic boundary conditions to ensure that the overlap is not computed
between correlated replicas. The resulting histograms are merged in an
accumulation operation similar to regular observables.

Improving scaling with process count will require a lower overhead
implementation of the spin overlap measurements---a problem we intend to
tackle in the near future.

\begin{figure*}[htbp]
\begin{center}
\includegraphics[width=\textwidth]{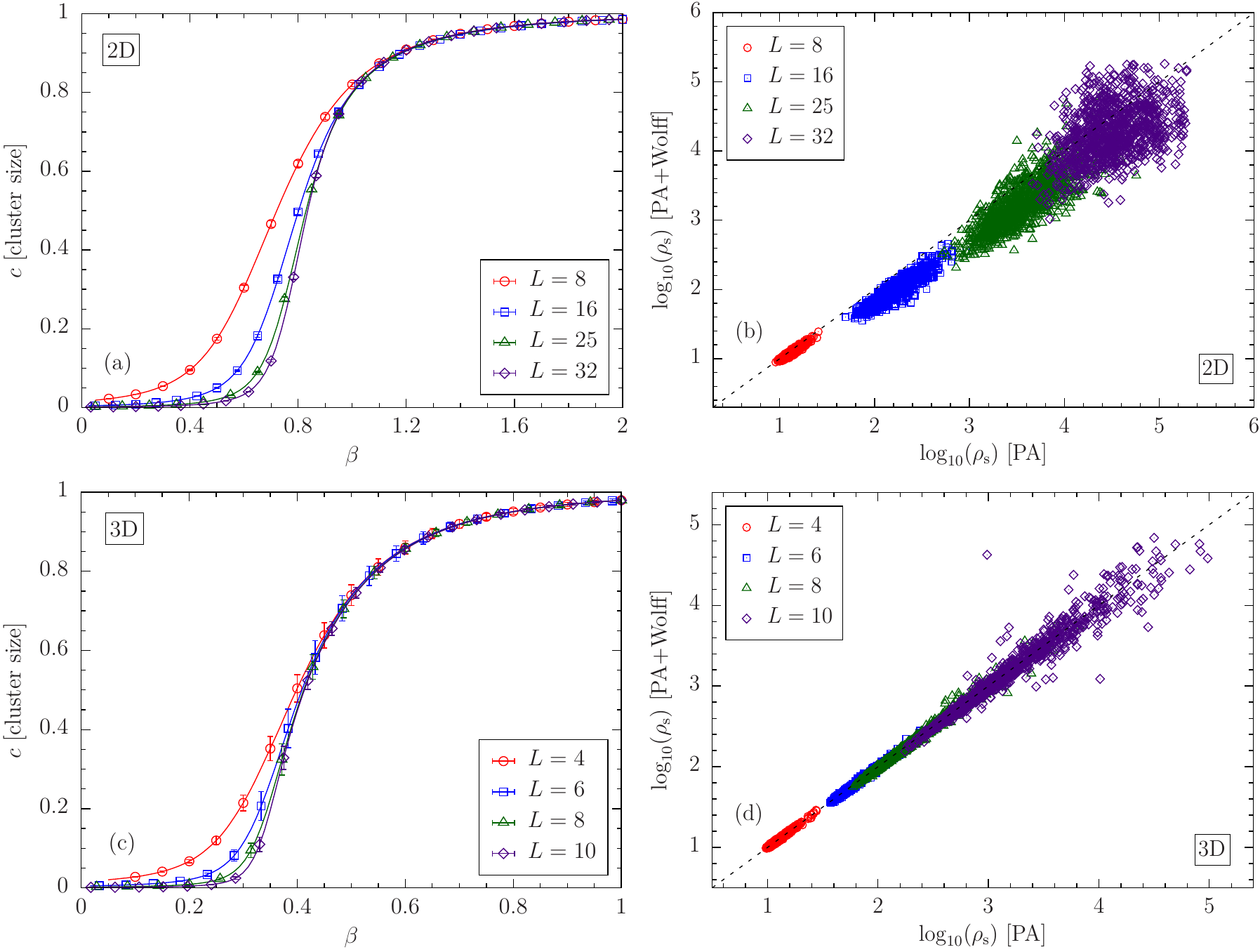}
\caption{
Mean normalized cluster size as a function of $\beta$ for the Wolff
algorithm [(a) and (c)] as well as the performance of the algorithm in
both 2D and 3D. There is marginal speed-up in 2D (b) and no discernible
speed-up in 3D (d).
}
\label{Wolff}
\end{center}
\end{figure*}

\section{Conclusions and future challenges}
\label{cc}

We have investigated various ways to optimize PAMC, ranging from
optimizations in the implementation, to the addition of accelerators, as
well as massively parallel implementations.  Many of these optimizations
lead to often considerable speed-ups. We do emphasize that these
approaches and even the ones that showed only marginal performance
improvements for spin glasses in 2D and 3D might be applied to other
approaches to simulate statistical physics problems potentially
generating sizable performance boosts. The reduction in thermal
error studied in this work can most directly be applied to the study
of spin glasses by providing more CPU time for disorder averaging.

For the study of spin glasses, our results show that the best
performance for PAMC is obtained by selecting the spins in a fixed
order, i.e., sequentially or from a checkerboard pattern. Similarly,
LBLT and TSPL schedules yield the best performance with LBLT having the
least parameters to tune and thus easier to implement. The number of
temperatures needed for annealing is remarkably robust for large system
sizes. Hence, in order to tackle hard instances, it is often convenient
to increase the number of sweeps rather than merely using more
temperatures. Dynamic population sizes are desirable, albeit at the cost
of a larger memory footprint. However, this can be easily mitigated via
massively parallel MPI implementations. In conjunction with
Ref.~\cite{amey:18}, and as far as we know, this study represents the
first analysis of PAMC from an implementation point of view.

Recently, we learned \cite{comment:mw} that the equilibration population
size $\rho_{\rm f}$ can be measured in a single run using a blocking
method. It would be interesting to further investigate and test this
idea thoroughly in the future. With an optimized PAMC implementation, it
would be interesting to also perform large-scale spin-glass simulations
to answer some of the unresolved problems in the field, such as the
nature of the spin-glass state in three and four dimensions. We plan to
address these problems in the near future.

\acknowledgments 

We thank Jonathan Machta and Martin Weigel for helpful discussions and
sharing their unpublished manuscripts.  H.~G.~K.~would like to thank
United Airlines for their hospitality during the last stages of this
manuscript.  We acknowledge support from the National Science
Foundation, NSF Grant No.~DMR-1151387.  The research is based upon work
supported in part by the Office of the Director of National Intelligence
(ODNI), Intelligence Advanced Research Projects Activity (IARPA), via
MIT Lincoln Laboratory Air Force Contract No.~FA8721-05-C-0002. The
views and conclusions contained herein are those of the authors and
should not be interpreted as necessarily representing the official
policies or endorsements, either expressed or implied, of ODNI, IARPA,
or the U.S.~Government.  The U.S.~Government is authorized to reproduce
and distribute reprints for Governmental purpose notwithstanding any
copyright annotation thereon. We thank Texas A\&M University for access
to their Ada and Curie HPC clusters. We also acknowledge the Texas
Advanced Computing Center (TACC) at The University of Texas at Austin
for providing HPC resources that have contributed to the research
results reported within this paper.

\appendix

\section{Wolff cluster updates}
\label{appendix:Wolff}

For the Wolff algorithm, we first measure the mean cluster size per
spin, as shown in Figs.~\ref{Wolff}(a) and \ref{Wolff}(c) for the 2D and
3D cases, respectively. Note the smooth transition of the mean cluster
size from $0$ to $1$. We identify a temperature range where the mean
cluster size is in the window $[0.1,0.9]$ \cite{comment:ranges}.  We
perform Wolff updates in this temperature range, i.e., we perform $10$
Wolff updates in addition to the $10$ regular Metropolis lattice sweeps
for each replica. The comparison of $\rho_{\rm s}$ with regular PAMC is
shown in Figs.~\ref{Wolff}(b) and \ref{Wolff}(d). While the Wolff
algorithm speeds up ferromagnetic Ising model simulations in 2D, the
speed-up is marginal for 2D spin glasses because of the zero-temperature
phase transition. In 3D, the Gaussian spin glass has a phase transition
near $T_{\rm c} \approx 0.96$, but the temperature window where the
Wolff algorithm is effective is much higher than $T_{\rm c}$. The
speed-up is therefore almost entirely eliminated, presumably because the
Metropolis algorithm is already sufficient for these high temperatures.
The fact that the Wolff algorithm is more efficient in 2D than 3D is
because clusters percolate faster in 3D, again rendering the effective
temperature range higher in 3D.  Therefore, Wolff updates constitute
unnecessary overhead in the simulation of spin glasses in conjunction
with PAMC.

Even though PAMC with the Wolff algorithm does not appear to work very
well for spin glasses, this does not mean they cannot be used together.
For example, in two-dimensional spin glasses, adding the Wolff algorithm
still has marginal benefits. The combination of PAMC and the Wolff
cluster updates can be used for ferromagnetic Ising models for the
purpose of parallel computing, because parallelizing the Wolff algorithm
while doable, is challenging. In population annealing, however, this can
be easily parallelized at the level of replicas, and not within the
Wolff algorithm itself.

\bibliographystyle{apsrevtitle}
\bibliography{refs,comments}

\begin{thebibliography}{45}
\bibitem[{com({\natexlab{c}})}]{comment:tmin}
\bibinfo{note}{{It is worth noting that optimizing $T_{\rm min}$ for various
  annealing schedules is not necessary because $\rho_{\rm s}$ is a
  monotonically-increasing function of the temperature. Thus, a higher $T_{\rm
  min}$ with the same number of temperature steps trivially results in a better
  thermalization.}}
\bibitem[{com({\natexlab{d}})}]{comment:pop}
\bibinfo{note}{{Note that the uniform population size is a special case of this
  generalized population size schedule.}}
\bibitem[{com({\natexlab{f}})}]{comment:ranges}
\bibinfo{note}{{We have also experimented with other ranges, such as
  $[0.2,0.8]$ or $[0.3,0.7]$. They consistently gave a similar or slightly
  worse speed-up.}}
\end{thebibliography}

\end{document}